# Improved Tactile Resonance Sensor for Robotic Assisted Surgery

David Oliva Uribe*[1], Johan Schoukens[1], Ralf Stroop[2]

Corresponding author*
[1]Vrije Universiteit Brussel, Brussels, Belgium
[2]Department of Stereotactic Neurosurgery, St. Barbara Hospital, Hamm, Germany

Keywords: Tactile sensor; Tissue differentiation; Piezoelectric Bimorph; Multisine excitation; Brain Tumour resection.

## Abstract:

This paper presents an improved tactile sensor using a piezoelectric bimorph able to differentiate soft materials with similar mechanical characteristics. The final aim is to develop intelligent surgical tools for brain tumour resection using integrated sensors in order to improve tissue tumour delineation and tissue differentiation. The bimorph sensor is driven using a random phase multisine and the properties of contact between the sensor's tip and a certain load are evaluated by means of the evaluation of the nonparametric FRF. An analysis of the nonlinear contributions is presented to show that the use of a linear model is feasible for the measurement conditions. A series of gelatine phantoms were tested. The tactile sensor is able to identify minimal differences in the consistency of the measured samples considering viscoelastic behaviour. A variance analysis was performed to evaluate the reliability of the sensors and to identify possible error sources due to inconsistencies in the preparation method of the phantoms. The results of the variance analysis are discussed showing that ability of the proposed tactile sensor to perform high quality measurements.

## 1 Introduction:

This contribution presents a tactile resonance sensor able to differentiate soft materials with similar mechanical characteristics. The intended application of this research work is to assist neurosurgeons by developing intraoperative tactile tools to improve tissue tumour delineation and tissue differentiation. The proposed solution introduces the use of a piezoelectric biomorph where mechanical vibrations are used to excite mechanically soft biological tissues or tissue phantoms by the use of multisines as excitation signal.

### *1.1 Problem*

The surgeon's tactile sense is – next to the visual aspect of the operative situs – crucial for the surgical procedure and strategy. Tactile perception of tissue consistency will be experienced by the surgeon either by direct manual palpation or transmitted by a surgical instrument in a tool-to-tissue interaction. This sensor modality, in combination with an eye-minded surgical technique, supports the delineation of tissue pathologies differing in consistency, as nodules, tissue maceration or tumour entities. Even different tissue compartments, textures, or surface structures of healthy tissue may be differentiated by direct or indirect tactile sense, ascribing it to a crucial influence for the surgical course.

With the introduction and the enhancements of the modern, so called less or minimal invasive surgical techniques, as laparoscopic surgery, neuroendoscopy [1], or more recently the NOTES (Natural Orifice Transluminal Endocsopic Surgery) technology [2], the surgeon's tactile sense is considerably constrained. For instance, due to the friction of the trocars at the entry ports at the abdominal or thoracic wall, transduction of tissue consistency is widely or even completely damped. Moreover, in robotic guided soft tissue surgery a tactile sensor modality is a priori not available and - if necessary - has therefore to be artificially emulated.

Reproducing the human sense of touch is a challenging task, due to the fact that human tactile sensing is a highly complex function composed of different sensory qualities. Next to nociceptors and thermoreceptors, a set of mechanoreceptors with different adapting rates enable the recognition of several object's properties like size, position, stiffness, vibration, texture, and roughness. Post processing of psychophysically information is important because it influences absolute sensory threshold, at which a stimulus becomes perceivable and it modulates the difference threshold, allowing discrimination of different stimulus strengths known as the just noticeable difference (JND). Furthermore, human sense of touch is subjected to habituation and adaptation [3].

Focussing on tissue consistency, a correct description of the mechanical behaviour of tissues could roughly be simulated by their elastic behaviour using a single parameter or element (e.g. purely elastic spring). For this task, more accurate models comprising viscoelastic properties should be used. These can be represented, for instance, by a standard viscoelastic solid model built up of two springs and a viscous damper element [4]. Even more, in order to provide a better description of the observed material behaviour on biological tissues, the implementation of hyper-viscoelastic constitutive models is frequently used, in particular for finite element calculations [5].

For preparatory and first experimental measurements, tissue phantoms are preferred to be used to mimic mechanical properties of biological tissues. The use of phantoms instead of real tissue presents several advantages, like a higher stability and the possibility to obtain compositions with controllable gradients [6]. The report of [7] compared the properties of tissue phantoms made of rubberlike hardened liquid plastics and porcine skin gelatine. For simulating tissue elasticity, individually, rubberlike hardened liquid plastics exhibit good characteristics and the capability of a long-term use. However, plastic phantoms lacked in a linear relation of concentration (i.e. the ratio of regular base liquid plastic to a softener solution) to their elastic modulus. Porcine skin gelatine gel preparations exhibit better tissue like characteristics where similar viscoelastic properties can be mimicked. Furthermore, it is reported in literature that identification of their mechanical parameters can be done using a discrete 1-D viscoelastic model [8].

*1.2. State of the art*

Several tactile sensor applications have been proposed for medical purposes. According to their type of application, these can be classified as: contact or non-contact sensors. With regard to their sensing principle, a proper classification can be done as follow: resistive, piezoelectric, capacitive, optical or resonant sensors. Furthermore, tactile sensors can be typified by their time resolution, their dynamical range or the spatial resolution and their construction principle (i.e. point or matrix array) [9]. A thorough review of the state of the art of tactile sensors and their applications can be found in [10,11].

Resonance driven tactile sensors based on piezoelectric transducers were first described by Omata [12], The sensing principle is based on the use of two piezoelectric transducers, one functioning as a vibration element and a second one for sensing. Different pathological tissue alterations as tumour, fibrosis and calcification were correlated using the evaluation of the frequency shift, $\Delta f = f_1 - f_2$, where $f_1$ is the non-contact resonance frequency and $f_2$ is the resonance frequency of the tactile sensor in contact with a tissue sample. Frequency shift evaluation reveals to be a stiffness-sensitive parameter, proofing a linear correlation of $\Delta f$ and the elasticity modulus of the sample under test. This sensing principle was lately implemented to develop several medical applications to determine tissue stiffness. It was used for different in vitro examination as in prostate tissue, for monitoring intraocular pressure, detection of lymph node metastasis, and skin lesions [13,14]. Nevertheless, the so far described strategy only provided a correlation based on tissue stiffness, hence its use on biological tissues with predominant viscoelastic behaviour (e.g. brain tissue) is limited.

*1.3. Improved sensing method*

In addition to the evaluation of tissue stiffness, damping estimation is necessary to provide further information to support decision making during surgical resection of evaluated tissue. In a proof of principle study, a more advanced measurement strategy was implemented by [15]. This approach used an analogue signal generation and a phase-locked loop circuit implemented in a computer software based tool. Differentiation of gelatine probes with minimal concentration differences was achieved using information related to changes in stiffness and damping. Additionally, different porcine ex vivo organs' tissues were tested using the piezoelectric bimorph sensor and a differentiation method through clustering of the measured data was proposed.

The same results could be achieved by the bimorph sensor when attached to a 6 DOF-robot equipped with a force-torque sensor to control the contact impression. Later an improved fully digital solution capable to perform measurements with high sensitivity for punctual tissue investigations was implemented for measuring coincidentally frequency shift and signal damping as a measure for tissue stiffness and damping behaviour [16].

For the sake of calibration and to assure reliability in the preparation of the samples, measurements of viscoelastic parameters by standard indentation methods have been performed [17]. Load-depth indentation, relaxation, and creep response tests were applied to identify absolute elasticity and viscosity constants. These measurements intend prospectively to establish a tissue consistency database of normal and pathological tissues [7].

In order to improve the measuring time and to avoid sensor-induced prolonged tissue deformation, a further enhancement was performed by driving the piezoelectric bimorph using multisine excitation to obtain the frequency response function (FRF) of the sensor in contact with a tissue sample [18]. The use of periodic excitations (i.e. multisine), simplified the calculation of the FRF by simple division of the output by the input spectra $G(j\omega_k)=Y_k/U_k$ [19]. In addition, for low levels of tissue deformation and quasi-instantaneous measurements, the use of identification techniques for linear systems can be used to obtain an accurate differentiation criterion, due to the fact that levels of nonlinearities are minimal and tissue's relaxation effect does not influence the measurement.

Moreover, the fast measurement time and fast signal processing capability of the bimorph tactile sensor system make it suitable for its implementation as a robotic surgical tool for assisted surgery for precise intraoperative real time diagnoses.

## 2. Materials and Methods:

### 2.1. *Tissue phantoms and ex vivo porcine tissue*

Biological soft tissues exhibit viscoelastic behaviour, therefore the preparation of phantoms was based on the use of a material with intrinsic relaxation and stiffness properties. Plastic rubber and silicone probes showed a minimal level of relaxation behaviour, therefore gelatine probes were used instead as a mimicking phantom. Gradually concentrated gelatine gel samples in the range from 10 % up to 16 % w/w (step size of 2%) have been prepared for the evaluation and characterisation using the piezoelectric tactile sensor according to [20].
In addition, fresh ex vivo porcine tissue probes delivered by a slaughterhouse were used to test the tactile sensor. Porcine brain hemispheres, brain slices and liver sections have been examined within 4 hours after the animals have been slaughtered.

### 2.2 Piezoelectric tactile sensor

The core element of the tactile sensor is a piezoelectric bimorph compound by two piezoelectric layers forming a cantilever beam. The tip is covered with a spherical plastic contact ball. One bimorph layer functions as an actuator and is excited by a broadband periodic signal $U_{actuator}$ to induce mechanical vibrations. The second layer is used as a sensing element, where the generated voltage $U_{sensor}$ is a function of the mechanical impedance of the whole system. For the purpose of this investigation, only random phase multisines were applied as excitation signal.

Although the bimorph sensor is not driven in resonance, the measurement of the FRF, calculated as the ratio of both bimorph's voltages, provides the necessary information to evaluate the resonance frequency and the maximum amplitude $U_{sensor}/U_{actuator}$, if a correct selection of the frequency components of the multisine is done (i.e. the bandwidth range covers the resonance frequency to evaluate). With the FRF information, a linear second order model is estimated using the algorithm ELiS (Estimator for Linear System) proposed by [21] and the pole zero plots are used for differentiation criterion to evaluate changes in the mechanical properties of the tissue or phantoms under test.

The tactile sensor, shown in Figure 1 can be mounted in a convenient housing jacket for remote manipulation using a robotic system. For measurements using the automated setup, the tactile sensor was mounted into a simple fixing structure. To avoid noise, the bimorph is electrically isolated from the housing and the fixing structure.

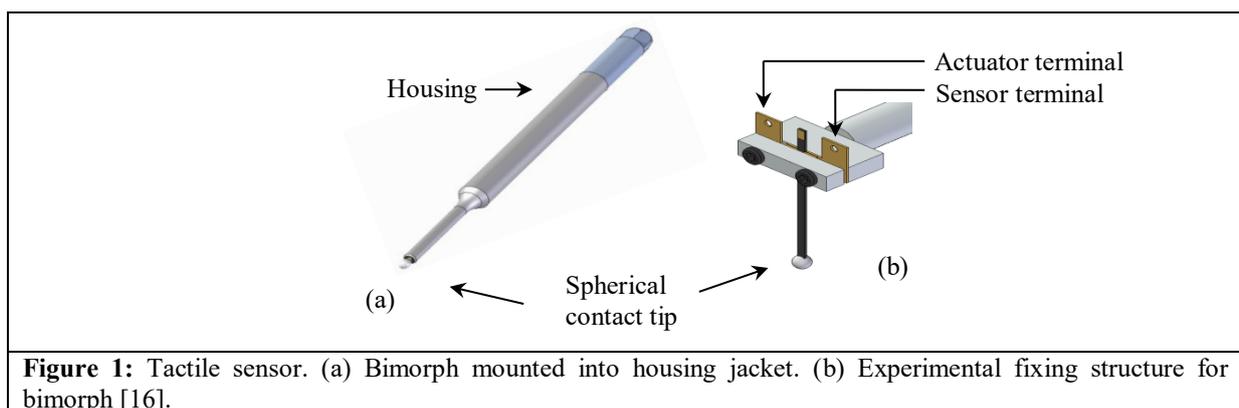

**Figure 1:** Tactile sensor. (a) Bimorph mounted into housing jacket. (b) Experimental fixing structure for bimorph [16].

## 2.3 Experiment setup

The automated measurement setup is shown in Figure 2. The tactile sensor with spherical tip is mounted, for vertical displacement, to a linear stage driven by a stepper motor with a resolution of 500 nanometers per step. Tissue or phantom samples can be placed on a second linear stage for horizontal displacement with a resolution of 10 micrometers per step. Both stages are controlled using a multipurpose data acquisition card and a Labview® application program. The contact force between the tactile sensor's tip and the sample is measured using a load cell and the displacement of the tip is obtained using a linear quadrature incremental encoder.

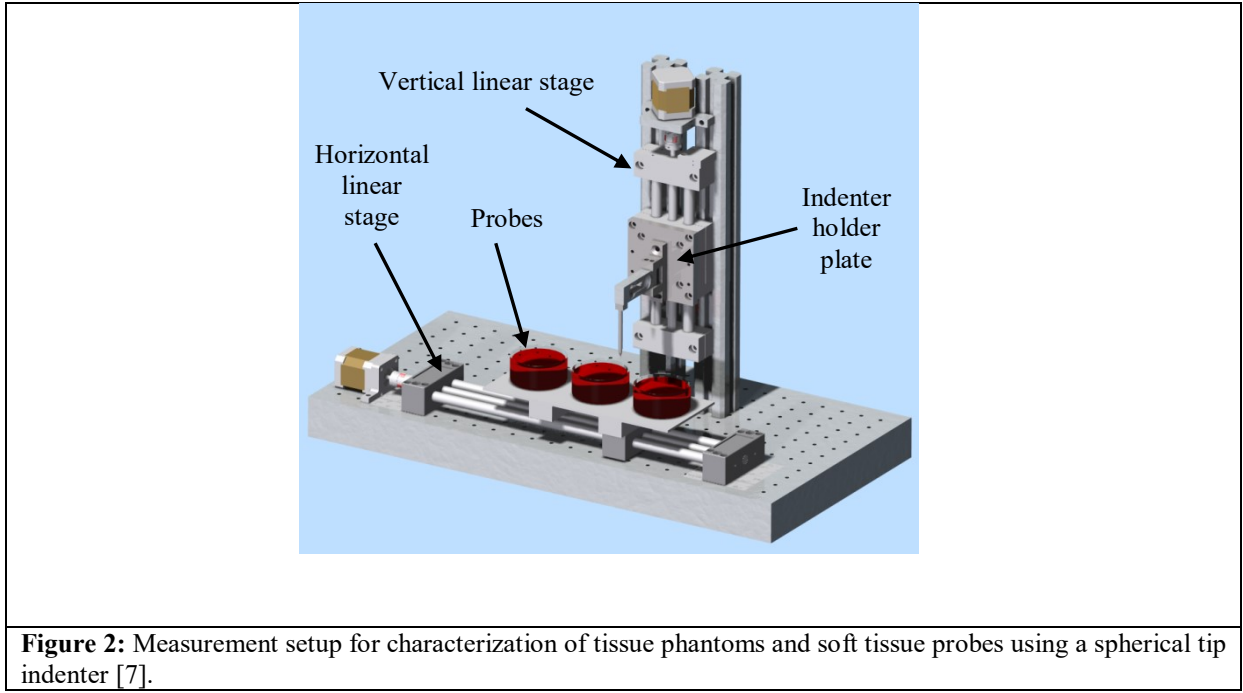

**Figure 2:** Measurement setup for characterization of tissue phantoms and soft tissue probes using a spherical tip indenter [7].

## 2.4 Multisine excitation for tactile sensor

In order to achieve fast measurements using the bimorph sensor, a broadband excitation signal was used. The selected signal is a random phase multisine. A multisine is a broadband periodic excitation signal with excellent properties to measure and identify linear and non-linear systems. It consists of a sum of harmonically related sinusoids [22]

$$u(t) = \sum_{n=1}^{N} A_n \{ e^{j(2\pi f_n t + \phi_n)} + e^{-j(2\pi f_n t + \phi_n)} \}$$

$$= \sum_{n=1}^{N} 2 A_n \cos(2\pi f_n t + \phi_n)$$

where N is the total number of sinewave components, $f_n$ are the excited frequency lines, $A_n$ is the set of amplitudes and $\varphi_n$ are the phases. One period of the broadband signal is defined by $1/\Delta_f$ where $\Delta_f$ is the frequency resolution. The total measurement time using multisine excitation $T_{bs}$ is defined as

$$T_{bs} = \frac{1}{f_0} + T_w$$

where $f_0 = \Delta_f$ and $T_w$ is the waiting time needed to let the transients to disappear. In this contribution, the frequency resolution is chosen to be 2.5 Hz with a frequency band from 10 Hz to10 kHz. In Figure 3 an example of the time domain signals for one period is depicted. The automated measurement setup generates the reference signal using an arbitrary waveform generator (AWG) and acquires the sensor signal $U_{sensor}$ and the actuator signal $U_{actuator}$ to calculate the FRF. In comparison to conventional step sine frequency sweep measurements, where each frequency line has to be excited individually and a waiting time to avoid transients has to be taken into account before measuring the input/output signals, the use of multisine excitation reduce considerably the measurement time, because all frequencies are excited in one period.

The period of the multisine is defined by the period of the lower excited frequency. The signal processing part is also performed rapidly as well, by the use of the Fast Local Polynomial Method [23], an advanced method to obtain the FRF accurately.
One important issue to take into account is that the total energy of the multisine is now dissipated over all excited frequency lines, and therefore a trade-off between sensitivity and input power has to be considered. Higher input power will increase the sensitivity of the tactile sensor, but the nonlinear contributions due to the bimorph itself have to be quantified to be sure they are not affecting the reliability of the measurements.

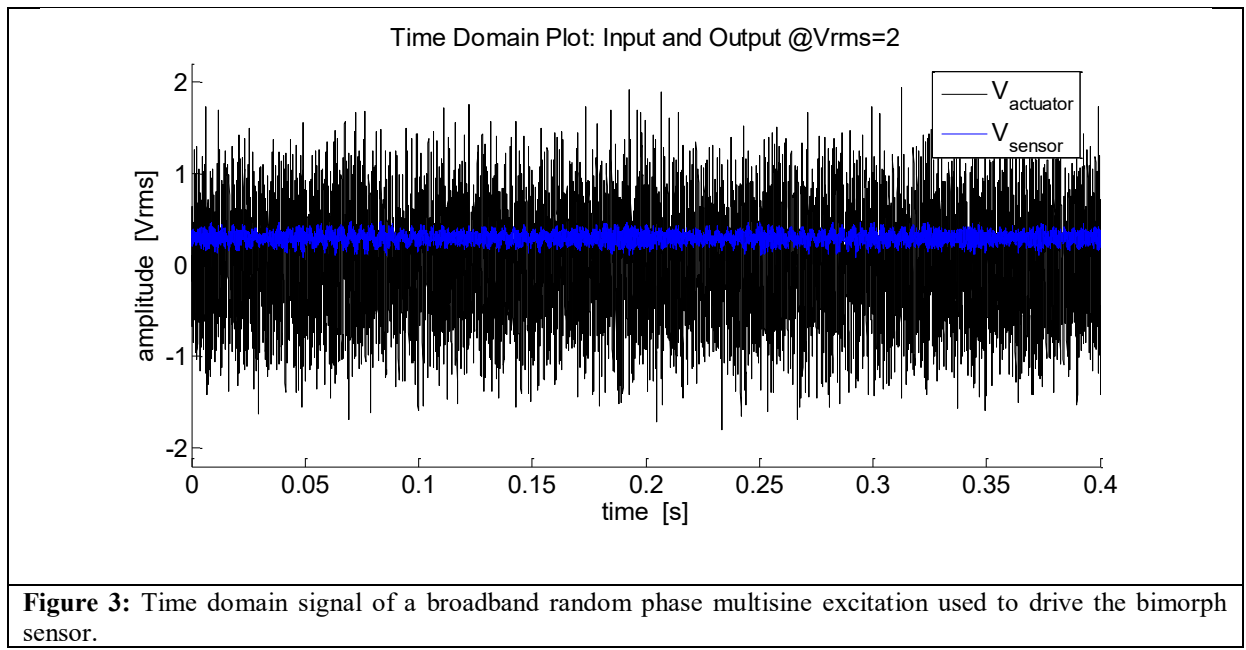

**Figure 3:** Time domain signal of a broadband random phase multisine excitation used to drive the bimorph sensor.

It is known that piezoelectric actuators and biological tissues exhibit nonlinear behaviour. To understand the influence of the level of the nonlinearities in the whole measurement procedure, a series of measurements of the input and output signals were performed at different input amplitudes in a range from 1 to 4 volts. As excitation signal was selected an odd random phase multisine with random harmonic grid Ngrid=4.

This choice is done to allow the analysis of nonlinear distortions [24]. The frequency lines of the excitation signal are divided in a block of four frequency odd lines where one is selected randomly as a detection line with amplitude equal to zero. Therefore, the output signal is composed by the measurement odd lines plus the contribution of the nonlinearities observed in the even and odd detection lines as seen in Figure 4.

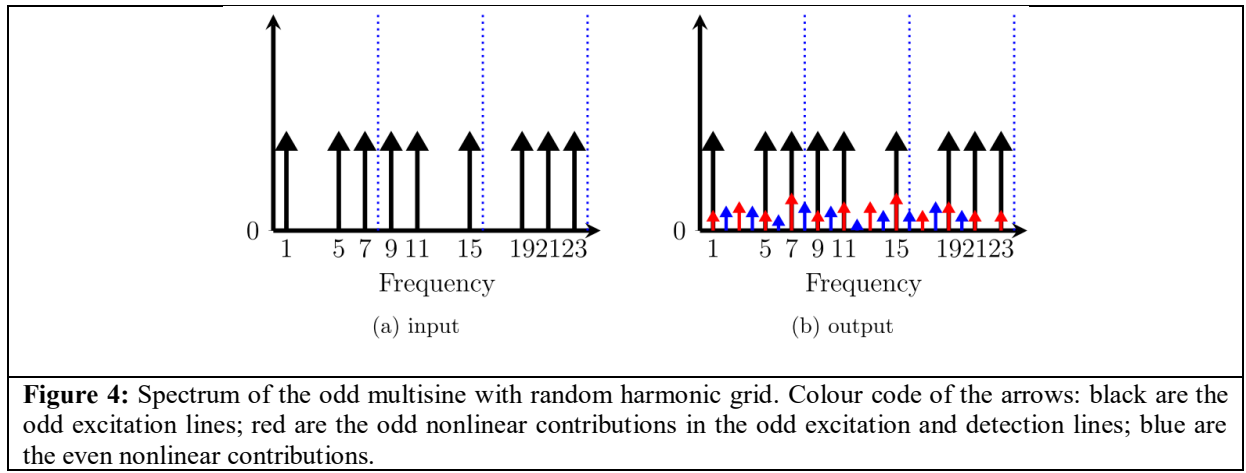

**Figure 4:** Spectrum of the odd multisine with random harmonic grid. Colour code of the arrows: black are the odd excitation lines; red are the odd nonlinear contributions in the odd excitation and detection lines; blue are the even nonlinear contributions.

It can be seen in Figure 5 that for a low-level input amplitude, the even and odd nonlinearities have almost no influence in the input and output signals, where the nonlinear contributions are -50dB lower than the amplitude of the measurement lines. For higher input amplitudes, the level of the odd nonlinear distortions increases in both input and output measurements. Around the neighbour frequencies that define the vibration modes, the presence of odd nonlinear distortions is even more evident. Although this behaviour, these contributions are -30dB lower than the amplitude in the measurement lines. In consequence, it is correct to assume that for this range, the sensor exhibit a quasi-linear behaviour.

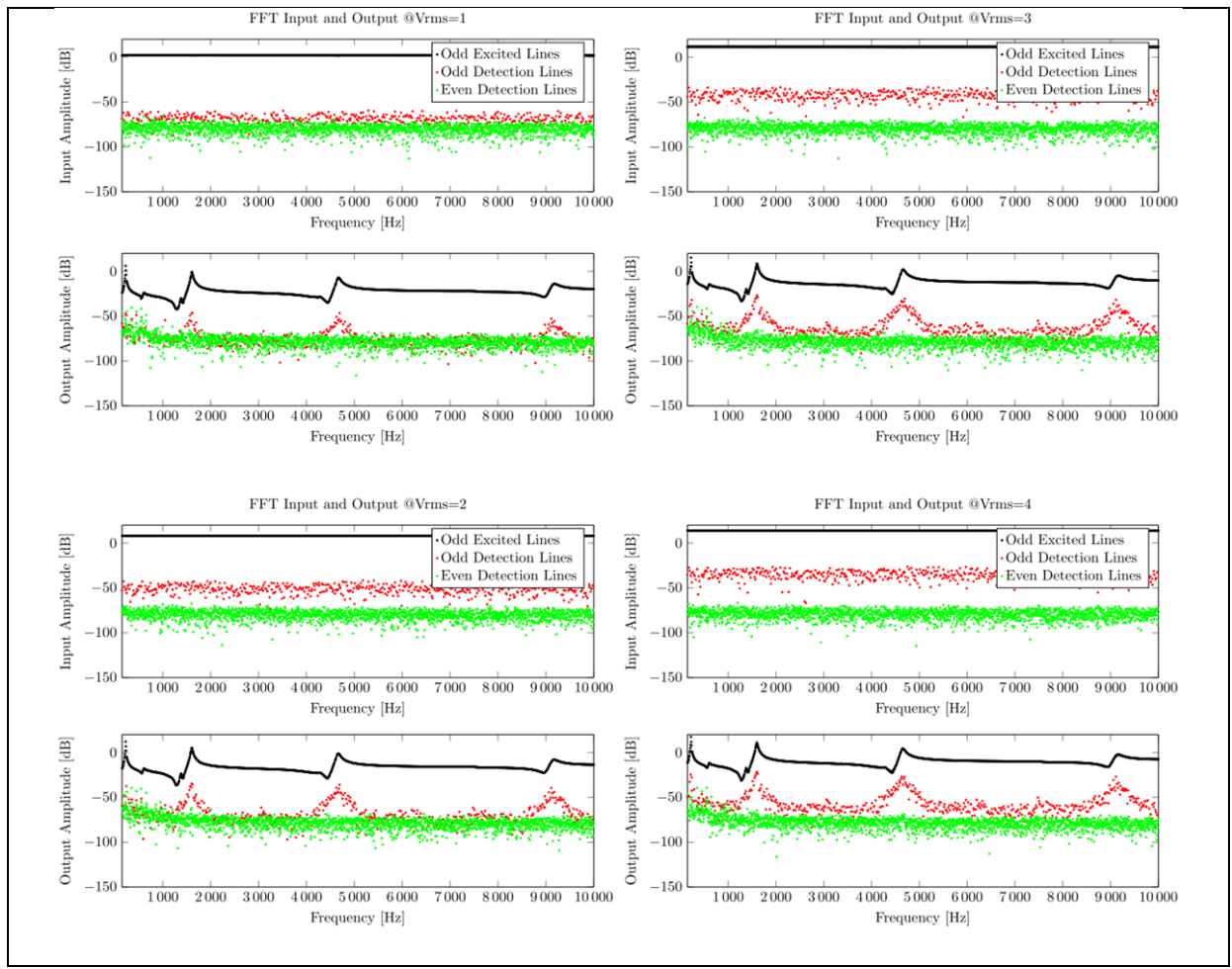

**Figure 5:** FFT of the input (top) and output (bottom) signals of the piezoelectric bimorph at 4 different levels of amplitude.

The selection of the frequency resolution is based on the evaluation of the second mode of the free vibrating sensor, which for the measurement of phantoms and ex vivo tissues is within the frequency band from 200 to 300 Hz. Figure 6 shows a zoom of the second mode of the free vibrating sensor of the bimorph when it is in contact with a gelatine phantom. The figure shows a comparison of the FRF measured at 4 different levels of input amplitude. This measurement shows that the selection of the frequency resolution is appropriate to obtain enough information around the resonance peak.

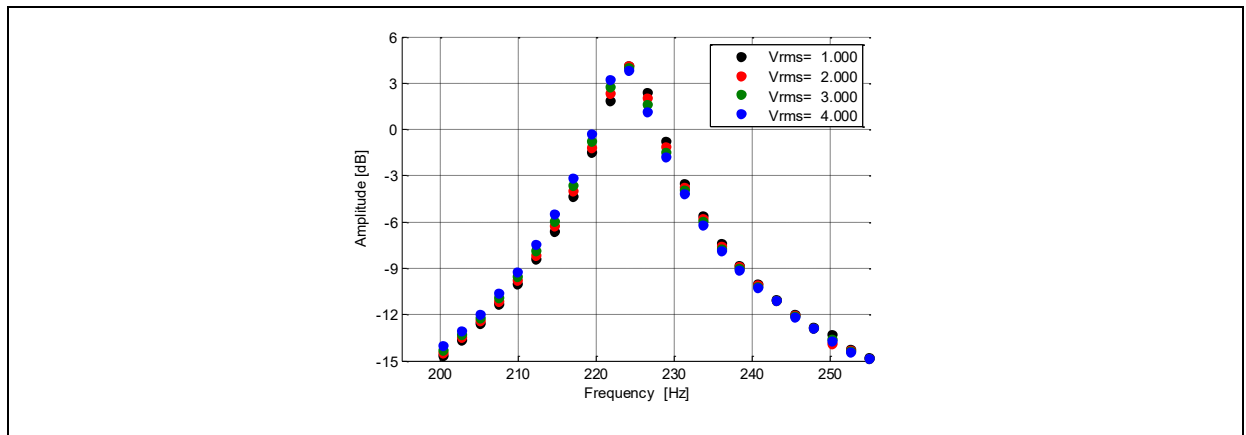

**Figure 6:** Comparison of the measurement of the FRF of the piezoelectric bimorph at 4 different levels of amplitude. Zoom of the second mode of the free vibrating sensor.

*2.5 Measurement protocols*

Viscoelastic relaxation effect is present in biological soft-tissues, as well as in solid matter, though to a much lesser degree. To observe the level of relaxation and time dependency of porcine ex vivo brain tissue, indentation tests using spherical tip indenter were conducted during a period of time of 180 seconds. Procedures for standard relaxation tests were programmed in the automated measurement setup. The acquired data were processed in Matlab® to obtain the material parameters using a viscoelastic solid model.

A series of gelatine phantoms probes with a gradient from 10 % to 16 % were utilised to evaluate the piezoelectric tactile sensor's sensitivity. The piezoelectric bimorph is driven using a random phase multisine excitation [25]. The piezoelectric bimorph is used as mechanical indenter. The automated measurement setup controls the indenter velocity and stops when the contact force is 1cN. After reaching the desired load, the measurement setup calculates a random phase multisine. The amplitude of the excitation signal is set to the maximum level of 4 volts in order to achieve the generation of mechanical vibrations at the highest amplitude. All performed measurements have a duration of 8 periods. The fast algorithm of the Local Polynomial Method [23] is used to process the information to estimate the nonparametric FRF. The frequency lines around neighbouring the first vibration mode are used to calculate the linear parametric estimation using the algorithm ELiS. The graphical representation of the position of the poles of the estimated linear second order model transfer function are used as a differentiation criteria. This graphical visualisation permits the user to identify possible changes in the mechanical properties of the samples in one or more measured regions. Therefore, the shifts in the poles' position can be related to changes in the viscoelastic properties.

The evaluation of the quality of the measurements is an important factor to determine the reliability of the presented piezoelectric tactile sensor. In this contribution, the analysis of the variance is used as a quality criterion. The mean and variance of the FRF of the bimorph in contact with a gelatine phantom was calculated to evaluate the variance of the FRF: 1 from one period to the other (over a total of eight periods); 2 over six repetitive measurements on the same point; 3 from one measurement point (position) to the other (over a total of six positions). A second phantom with the same concentration was prepared on a different day to obtain the FRF mean and variances above described. Finally, the difference of the FRF mean of the two samples was calculated to find possible errors due to variability in the sample preparation. For the variance analysis, the methods described in [23] were used.

## 3 Results:

### 3.1 Effect of fast relaxation in brain tissue and liver

The relaxation response test is performed by rapidly indenting the surface's probe up to a defined depth which then is kept constant for a defined interval of time and the contact force is measured. For viscoelastic materials (e.g. biological tissues) a time dependent load decay of the initial applied impact force is observed, known as the relaxation effect.

Measurements of the relaxation response on porcine ex vivo brain cortex, white matter and liver were conducted during 180 seconds with an initial contact force of 1 cN. As shown in Figure 7, all tissue samples tested presented a nonlinear viscoelastic relaxation, where in the first seconds of the measurement, the decay slope of the contact force is very fast and the initial contact force is reduced by at least 60%. This effect establishes the requirement to perform soft tissue consistency measurements with tactile sensors using a very fast measuring technique in order to avoid errors due to dependency of the output signal and the contact force between the sensor and the sample under test.

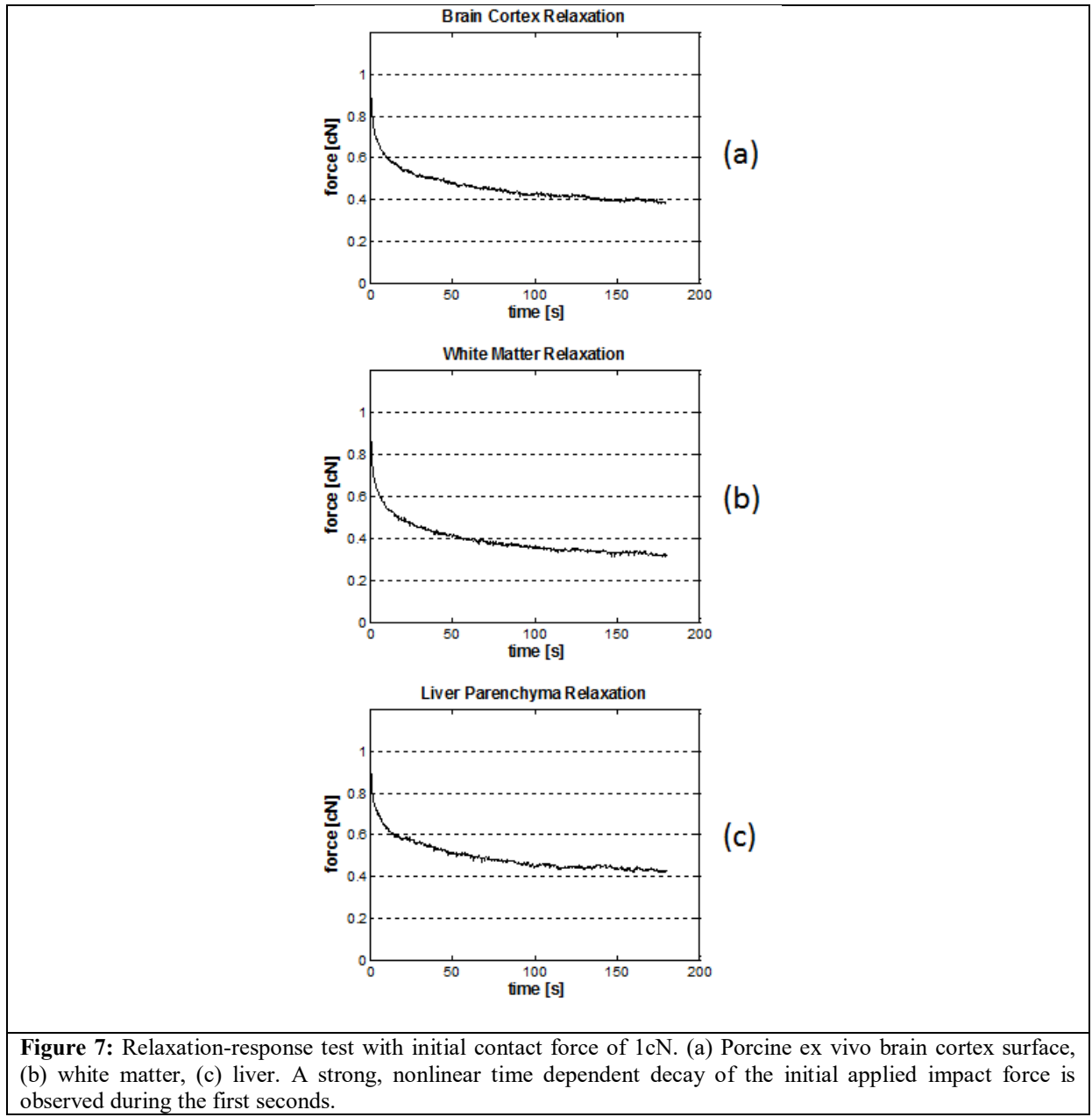

**Figure 7:** Relaxation-response test with initial contact force of 1cN. (a) Porcine ex vivo brain cortex surface, (b) white matter, (c) liver. A strong, nonlinear time dependent decay of the initial applied impact force is observed during the first seconds.

*3.2 Differentiation of gradually concentrated tissue phantoms*

A set of gradually concentrated gelatine phantoms of comparable consistency has been used to mimic mechanical brain tissue properties. In Uribe et al. 2009 the effectiveness of the bimorph to differentiate these phantoms was proved. Nevertheless, the obtained results were using a step sine sweep excitation, thus the measurement time was considerably large.

The gelatine gel concentrations were chosen to approximately cover the range of brain tissue as it was felt by a neurosurgeon. A difference (step) of 2 % in concentration gradation could not be distinguished by neurosurgeons, neither by manipulation with surgical instruments or by direct digital palpation.

Applying a multisine excitation signal, a series of gelatine phantoms with a gradient from 10% to 16 % in concentration were evaluated. The contact force between the sensor's tip and the surface's phantom was 1 cN. Figure 8a shows a plot of the FRF measurements of all phantoms. As seen in this plot, several resonance modes were excited along the frequency range. In order to be able to differentiate the phantoms it is necessary to zoom in the neighbourhood of the resonance peaks. The second mode of the free vibrating sensor exhibited the best sensitivity to detect changes in the concentration of the phantom (i.e. change in mechanical properties) as depicted in 8b. The graphical representation of the poles of the estimated second order model using ELiS is presented in 8c.

For this vibration mode, the bimorph's resonance frequency shifts upwards with increasing gelatine gel concentration, whereas the amplitude will be more and more damped. As seen here, using the piezoelectric tactile sensor, even small concentration differences could, without any difficulty, be distinguished. The measuring time for a single point takes approximately 400 ms per measured period. In practice, the complete evaluation of a point on a phantom can be done in less than 1 second since only 2 periods are required.

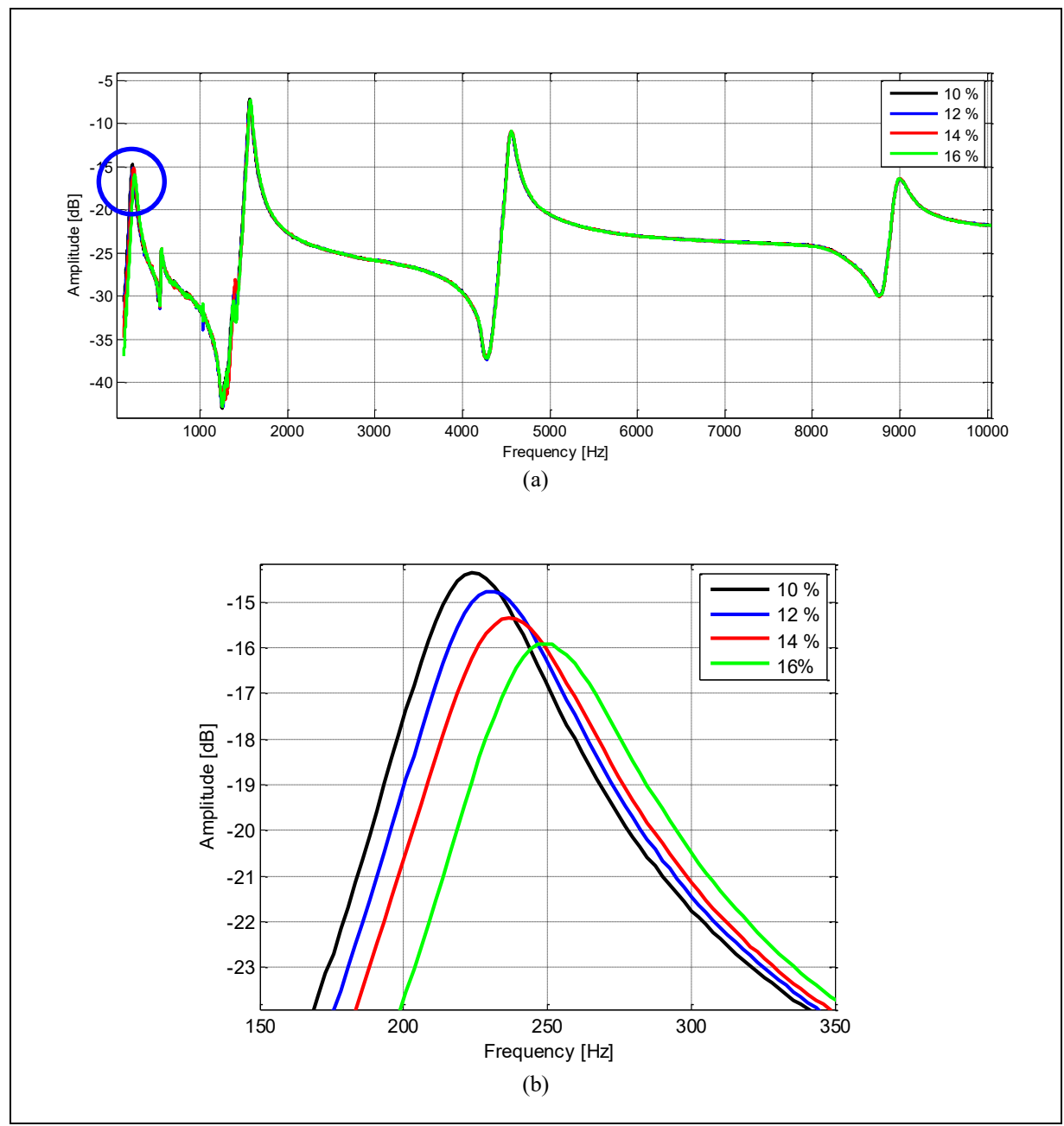

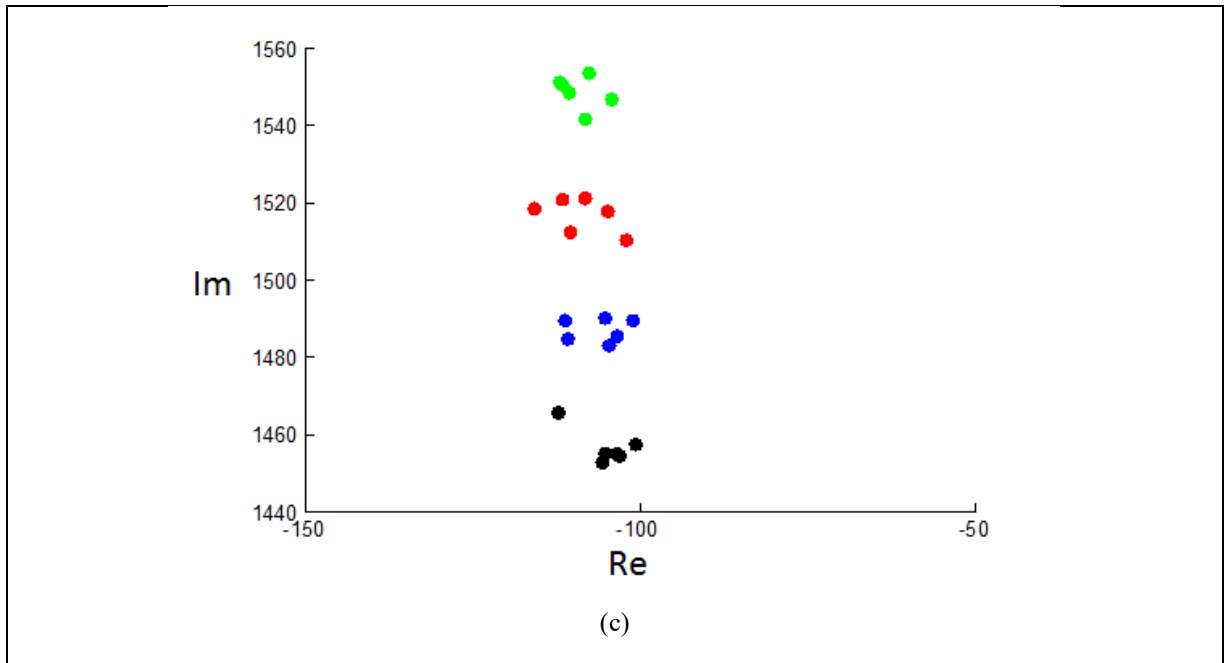

(c)

**Figure 8:** Broadband multisine excitation signal used for driving the bimorph. (a) FRF of the ratio $U_{sensor}/U_{actuator}$, (b) zoom to the second mode of the free vibrating sensor. (c) position of the poles of a second order model identified using ELiS for a series of four gelatine phantoms. Each pole represents the measurement of 1 point on the phantom using only 8 periods. Frequency resolution= 2.5 Hz

*3.3 Variance analysis*

The results of the variance analysis are presented in Figure 9. The variability from one period to the other is very small. This information is relevant to understand that under the presented measurement conditions, viscoelastic relaxation effect has a minimal influence.

The variance over repetition is higher than the one over periods, but still on a low level, showing that for a low level of contact force (e.g. 1 cN) the indentation produced by the bimorph did not cause severe plastic deformations. It is important to emphasise that for each repetition, a new realization of the excitation signal is made in order to reveal the presence of nonlinear distortions [26].

The low level of the variance over different positions shows that the phantom preparation was performed with high quality and homogeneity in their properties. The difference of the FRF means of the two preparations of the same concentration of the gelatine phantom exhibits the highest level, but still is lower -30 dB than the mean of the FRF, confirming that the preparation procedure of the phantoms is highly controlled to perform measurements with high repeatability. Moreover, the three calculated variances (i.e. over periods, repetitions and positions) confirm that the bimorph sensor can perform measurement with high precision and reliability.

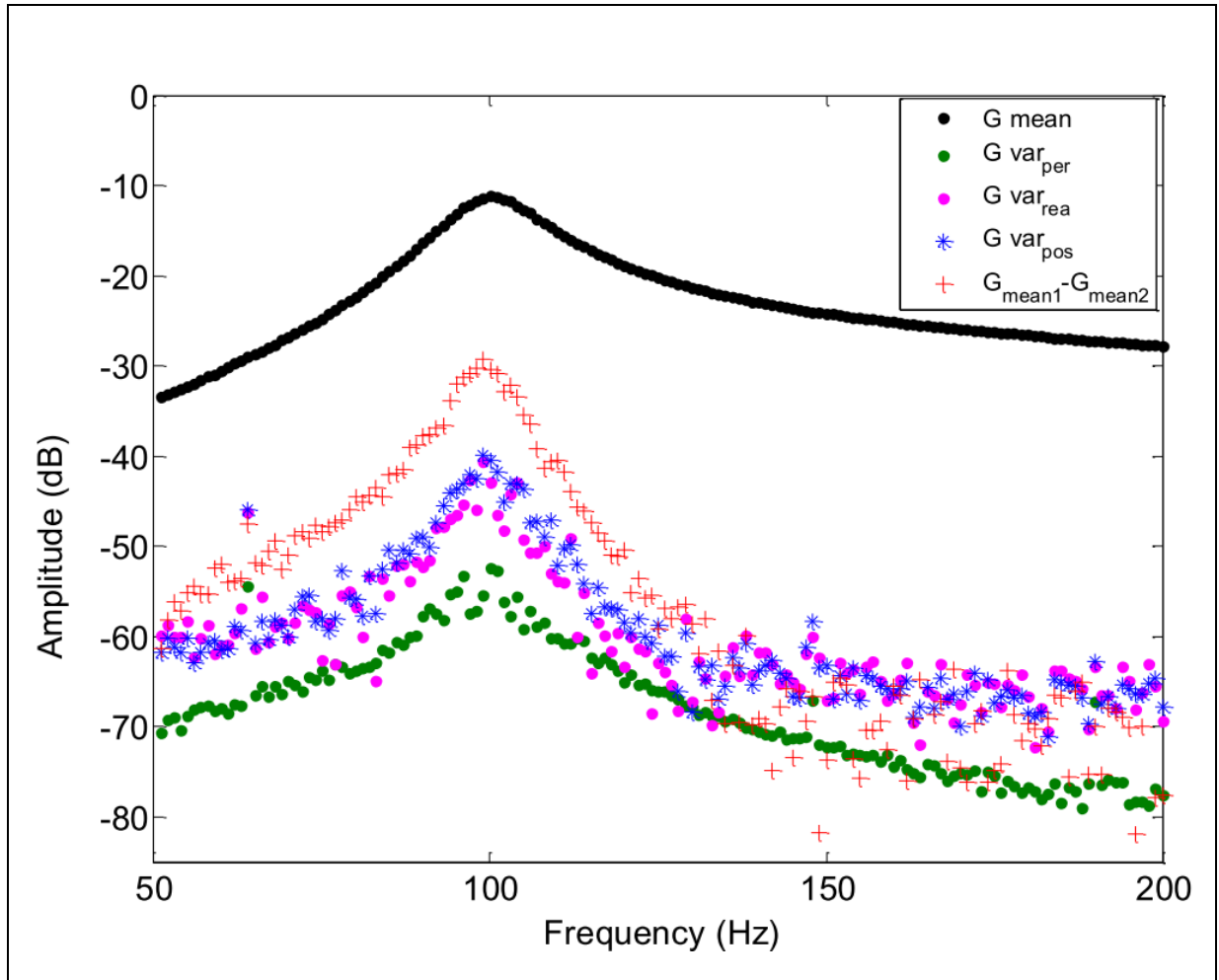

**Figure 9:** Variance analysis of the FRF of the ratio $U_{sensor}/U_{actuator}$. Black: FRF mean of the sample 1. Green: variance of the sample 1 over 8 periods. Magenta: variance of the sample 1 over 6 different repetitions on the same position. Blue: variance of the sample over 6 different positions. Red: difference of the FRF mean of sample 1 and sample 2.

## 4 Discussion:

The potential usage of piezoelectric tactile sensors in medical applications has already been reported in scientific literature. However, the focus has been set on elasticity determination [13,14]. The indentation measurements performed on ex vivo porcine brain and liver tissue showed that decaying and time depending viscoelastic effects are evident and cannot be disregarded.

Tissue relaxation effects gave consequently reason for the development of rapid measurement techniques. Furthermore, inherent low consistency contrast distribution, as it is expected for brain tissue, requires a high sensor resolution.

Therefore, we reported our efforts to provide an advanced solution using a tactile piezoelectric bimorph sensor using random phase multisine as excitation signal. The proposed technology provides a fast scanning device with high resolution to detect tissue consistency by the evaluation of its mechanical properties (i.e. viscoelastic properties), allowing discrimination of subtle differences in tissue phantom consistency. The measurement time is reduced significantly to less than 1 second per measurement point, and therefore, the result is provided in real-time without the influence of viscoelastic relaxation.

Relaxation is causally explained by a time dependent micro-shift of tissue water content due to the even moderate strain of 1 cN, caused by the sensor tip, and therefore altering tissue consistency. Relaxation processes are induced during every intraoperative tissue manipulation. Though, this process is in principle reversible, it is time-delayed. Shortly repeated measurements will therefore lead to slightly different results in tissue consistency.

The variance analysis showed that the bimorph sensor can performed measurements with high reliability, confirming that the level of nonlinear distortions for the proposed measurement conditions has a minimal influence and a linear approach can be used.

## 5 Conclusions:

This contribution presented a tactile sensor measurement system. The sensing element is a piezoelectric bimorph driven in self-sensing configuration. Using multisines as excitation signal, the bimorph vibrates and the properties of contact between the sensor's tip and a certain load are evaluated. The nonparametric FRF of the bimorph's signals $U_{sen}/U_{act}$ is calculated using the fast algorithm of the Local Polynomial Method.

A linear parametric second order model of the transfer function $U_{sen}/U_{act}$ is identified using ELiS, an iterative weighted least square estimator in the frequency domain. The distribution of the poles during tissue scanning is utilised as a criterion to identify changes in the properties, assuming a viscoelastic behaviour, of the measured samples. The ability of the tactile sensor to identify minimal differences in the consistency of soft materials was tested using gelatine phantoms. The variance analysis showed the reliability of the system and the protocol used for the preparation of the samples.

## Acknowledgments:

This work was founded by the Methusalem grant of the Flemish Government (METH-1), the Fund for Scientific Research (FWO), the IAP VI/4 DYSCO program and the European Research Council (ERC) Advanced ERC grant No 320378 – SNLSID (Johan Schoukens). David Oliva Uribe would like to thank the support of the Mexican National Council on Science and Technology (CONACYT).